\documentclass[twocolumn,showpacs,preprintnumbers,amsmath,amssymb]{revtex4}

\usepackage{graphicx}
\usepackage{dcolumn}
\usepackage{bm}
\usepackage{multirow} 

\begin{document}

\preprint{APS/KTT version 1.0}

\title{Isotope shifts of the (3s3p)$^3$P$_{0,1,2}$  - (3s4s)$^3$S$_1$  Mg I transitions}

\author{Ming He, Kasper T. Therkildsen}\email{kaspertt@fys.ku.dk}
\author{Brian B. Jensen, Anders Brusch}
\author{Jan W. Thomsen}

\affiliation{The Niels Bohr Institute, Universitetsparken 5, 2100
Copenhagen, Denmark}

\author{Sergey G. Porsev\footnote{Present address:
School of Physics, University of New South Wales,
Sydney, NSW 2052, Australia.}}
\affiliation{Petersburg Nuclear Physics
Institute, Gatchina, Leningrad district 188300, Russia}

\date{\today}

\begin{abstract}
We report measurements of the isotope shifts of the
(3s3p)$^3$P$_{0,1,2}$ - (3s4s)$^3$S$_1$ Mg I transitions for the
stable isotopes $^{24}$Mg (I=0), $^{25}$Mg (I=5/2) and $^{26}$Mg
(I=0). Furthermore the $^{25}$Mg $^3$S$_1$ hyperfine coefficient
A($^3$S$_1$) = (-321.6 $\pm$ 1.5) MHz is extracted and found to be
in excellent agreement with state-of-the-art theoretical predictions
giving A($^3$S$_1$) = -325 MHz and B($^3$S$_1$) $\simeq 10^{-5}$
MHz. Compared to previous measurements, the data presented in this
work is improved up to a factor of ten.

\end{abstract}

\pacs{Valid PACS appear here}
\maketitle

\section{\label{Intro}Introduction}

Accurate measurements of atomic transitions play an important role
in many parts of physics and astronomy. They form a basis for
state-of-the-art atomic structure calculations and provide important
reference data for spectroscopic measurements. Accurate
spectroscopic measurements have a wide range of applications
covering astrophysical \cite{Astro1} and laboratory based
experiments including optical cooling schemes and atomic clocks
\cite{Clock1,Clock2}. With the access to new and improved
spectroscopic data detailed models using relativistic many-body
methods of atomic structure calculations and models may advance
significantly \cite{calculation1,calculation2}. For the
alkaline earth elements some transitions are very well known, but
most transitions are relatively unknown or known only with a modest
accuracy.

The magnesium atom is particularly interesting in connection with
star evolution and for accurate spectroscopic measurements on
distant quasars in the search for a possible temporal drift of the
fine structure constant $\alpha =e^{2}/\hbar c$. For star evolution
models the (3s3p)$^3$P$_{0,1,2}$ - (3s4s)$^3$S$_1$ transitions play
an important role and here new data on the isotope shifts will be
helpful \cite{Astro1}. Absorption measurements from distant quasars
rely on accurate laboratory values of transition wavelengths and
isotope ratios.

In this Brief Report we present new improved measurements for the
isotope shift and hyperfine structure splitting of the
(3s3p)$^3$P$_{0,1,2}$ - (3s4s)$^3$S$_1$ Mg I transitions around 517
nm in a metastable atomic magnesium beam. We improve previous
measurements by up to a factor of ten for the stable isotopes
$^{24}$Mg (I=0), $^{25}$Mg (I=5/2) and $^{26}$Mg (I=0). For
$^{25}$Mg $^3$S$_1$ we extract the hyperfine coefficient
A($^3$S$_1$) and compare the result to state-of-the-art relativistic
many-body calculations.

\section{\label{Experimental setup}Experimental setup}

\begin{figure}[htb!]
\includegraphics*[width=1\columnwidth]{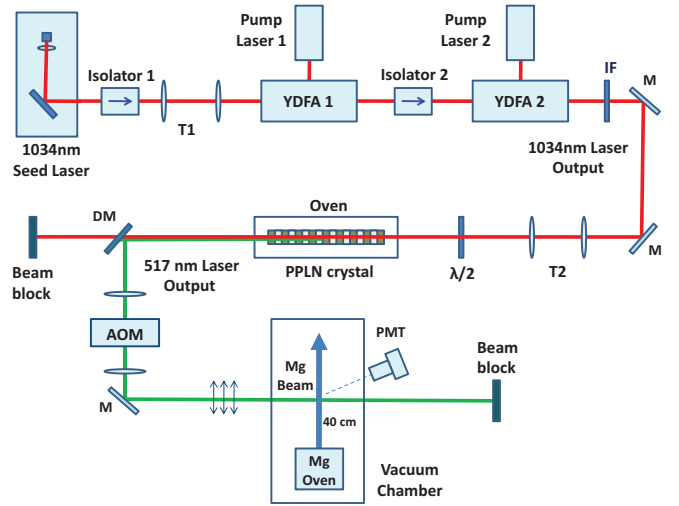}
\caption{(color online) Schematic diagram of the laser system: T,
telescope; DM, dichroic mirror; IF, interference filter; M, mirror;
$\lambda/2$, half waveplates; BS, beam splitter; AOM, acousto-optic
modulator; PMT, photomultiplier tube. A two-stage YDFA system is
used for 1034 nm laser amplification, and is single pass frequency
doubled in a PPLN crystal. Fluorescence spectroscopy is performed on
a metastable magnesium beam, approximately 40 cm from the oven
orifice.}\label{fig: Experimental setup}
\end{figure}

Fig.~\ref{fig: Experimental setup} shows the experimental setup
used for spectroscopy on metastable magnesium atoms. The Mg oven is
operated around 520 $^{\circ}$C and produces an effusive magnesium
beam with mean velocity 1000 m/s and flux of $\sim$10$^{13}$
atoms/s. Electron impact initiates the discharge which runs at a
stable current of about one ampere in a setup similar to the one
described in \cite{Mg metastable oven}. Using the main 285.3 nm
(3s$^2$)$^1$S$_{0}$ - (3s3p)$^1$P$_1$ transition we estimate about
40\% of the atoms are in a metastable state, distributed among the
$^3$P$_{0,1,2}$ levels. The 517 nm light is produced from a fiber
amplified diode laser centered at 1034 nm. We use an external cavity
diode laser in a Littrow configuration followed by a 40 dB optical
isolator. Typically the output after the isolator is 15 mW and about
10 mW is injected into a two-stage Yb-doped fiber amplifier (YDFA)
system. The fiber consists of a highly doped single-mode core of 10
$\mu$m diameter with a peak absorbtion of 1200 dB/m at 976 nm and a
large multimode pump guiding cladding of 125 $\mu$m in diameter. The
fibers are pumped with up to 5 W at 976 nm. After the amplifier
stage 1.5 W of 1034 nm light is generated. The 1034 nm light is
single pass frequency doubled in a periodically poled lithium
niobate (PPLN) crystal generating up to 40 mW at 517 nm. The domain
period of the PPLN crystal is 6.37 $\mu$m with a quasi-phase
matching temperature around 35$^{\circ}$C and temperature
coefficient of 0.09 nm/K. An oven is used to stabilize the PPLN
temperature within 0.01$^{\circ}$C in order to achieve optimal phase
matching. After the frequency doubling, a dichroic mirror is used to
separate the 517 nm light from the 1034 nm light.

Spectroscopy is performed 40 cm from the oven orifice using linearly
polarized light. The imaging system collects fluorescence from an
area of about 8 mm$^2$ within the atomic beam limiting the residual
Doppler effect to 60 MHz. A 275 MHz - 400 MHz AOM is used for
frequency scale calibration. The absolute AOM frequency is
controlled below 1 kHz RMS and was verified using a precision
counter. For calibration both zero and first order beams from the
AOM are overlapped producing a double set of spectra. Changing the
dc offset of the AOM enable us to test the degree of linearity of
the frequency scan. The intrinsic linewidth of the 517 nm light has
been measured to be below 3 MHz using a high finesse cavity. Each
spectrum is averaged 32 times and 30 different spectra are recorded
for each transition.

\section{Method of calculation}
\label{MetCalc}

The method used for calculation of the magnetic dipole and electric
quadrupole hyperfine structure constants $A$ and $B$ for the
$^3\!S_1$ state is a combination of the configuration interaction
(CI) method with many-body perturbation theory
(MBPT)~\cite{DzuFlaKoz96b}. Initially the method was developed for
calculating energy levels. The MBPT was used to construct an
effective Hamiltonian for valence electrons. Then the multiparticle
Schr\"odinger equation for valence electrons was solved in the frame
of the CI method. Following the earlier works, we refer to this
approach as the CI+MBPT formalism.

In this approach, the energies and wave functions are determined
from the equation
$$H_{\rm eff}(E_n) \Phi_n = E_n \Phi_n,$$
where the effective Hamiltonian is defined as
$$H_{\rm eff}(E) = H_{\rm FC} + \Sigma(E).$$
Here $H_{\rm FC}$ is the Hamiltonian in the frozen core
approximation and $\Sigma$ is the energy-dependent correction, which
takes into account virtual core excitations.

In order to calculate other atomic observables, one needs to
construct the corresponding effective operators for valence
electrons \cite{DzuKozPor98,PorRakKoz99P,PorRakKoz99J}. In
particular, the effective operator of the hyperfine interaction used
in this work accounts for the core-valence and core-core
correlations. To account for shielding of an externally applied
field by core electrons we have solved random-phase approximation
(RPA) equations, summing a certain sequence of many-body diagrams to
all orders of MBPT~\cite{DzuKozPor98,KolJohSho82}.

We consider Mg as a two-electron atom with the core
[1$s^2$,...,2$p^6$]. On the whole the one-electron basis set for Mg
consists of 1$s$--13$s$, 2$p$--13$p$, 3$d$--12$d$, and 4$f$--11$f$
orbitals, where the core- and 3,4$s$, 3,4$p$, 3,4$d$, and 4$f$
orbitals are Dirac-Hartree-Fock (DHF) ones, while all the rest are
the virtual orbitals. The orbitals 1$s$--3$s$ are constructed by
solving the DHF equations in V$^N$ approximation, 3$p$ orbital is
obtained in V$^{N-1}$ approximation, and 4$s$, 4$p$, 3,4$d$, and
4$f$ orbitals are constructed in V$^{N-2}$ approximation. We
determined virtual orbitals using a recurrent procedure similar to
Ref.~\cite{BogZuk83} and described in detail
in~\cite{PorRakKoz99P,PorRakKoz99J}. Configuration-interaction
states were formed using this one-particle basis set which is
sufficiently large to obtain numerically converged CI results. An
extended basis set, used at the stage of MBPT calculations, included
1$s$--19$s$, 2$p$--19$p$, 3$d$--18$d$, 4$f$--15$f$, and 5$g$--11$g$
orbitals.

The results obtained in the frame of the CI+MBPT method for the hfs
constants $A$ and $B$ are $A(^3\!S_1)$ = -325 MHz and $B(^3\!S_1)
\simeq 10^{-5}$ MHz. The theoretical value for the magnetic dipole
constant $A$ is in a good agreement with the experimental value
obtained in this work $A(^3\!S_1)$ = -321.6 $\pm$ 1.5 MHz.

The electric quadrupole hfs constant $B(^3\!S_1)$ is very close to
zero. In the absence of configuration interaction of the $3s4s$
configuration with other configurations (like $3s\,nd$ and $nd^2$
configurations, where $n \geq 3$) the hfs constant $B$ would be
exactly equal to zero. But (very weak) configuration interaction
leads to a non-zero value of $B$, though very small.

\section{\label{Results and discussion}Results and discussion}

\begin{figure}[htb!]
\centering
\includegraphics*[width=1 \columnwidth]{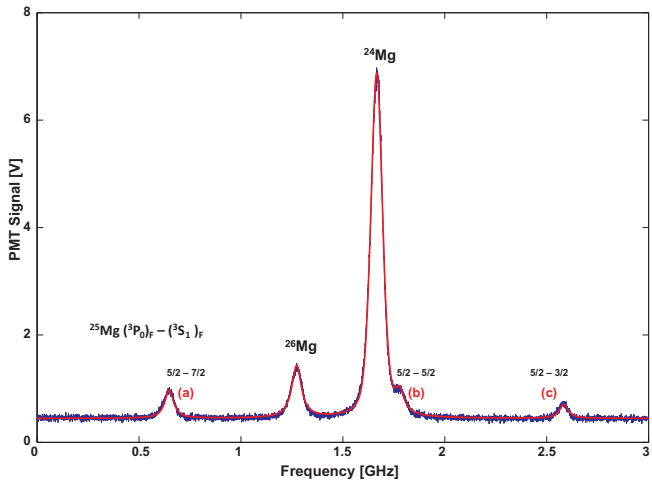}
\caption{(color online) Fluorescence signal from the $^3$P$_0$ -
$^3$S$_1$ transition. The blue curve represents the experimental
data and the red curve is a fit to the data. The hyperfine splitting
of $^{25}$Mg($^3$P$_0$)$_F$ - ($^3$S$_1$)$_F$ are indicated as (a) -
(c).}\label{fig: 3p0}
\end{figure}

\begin{figure}[htb!]
\centering
\includegraphics*[width=1\columnwidth]{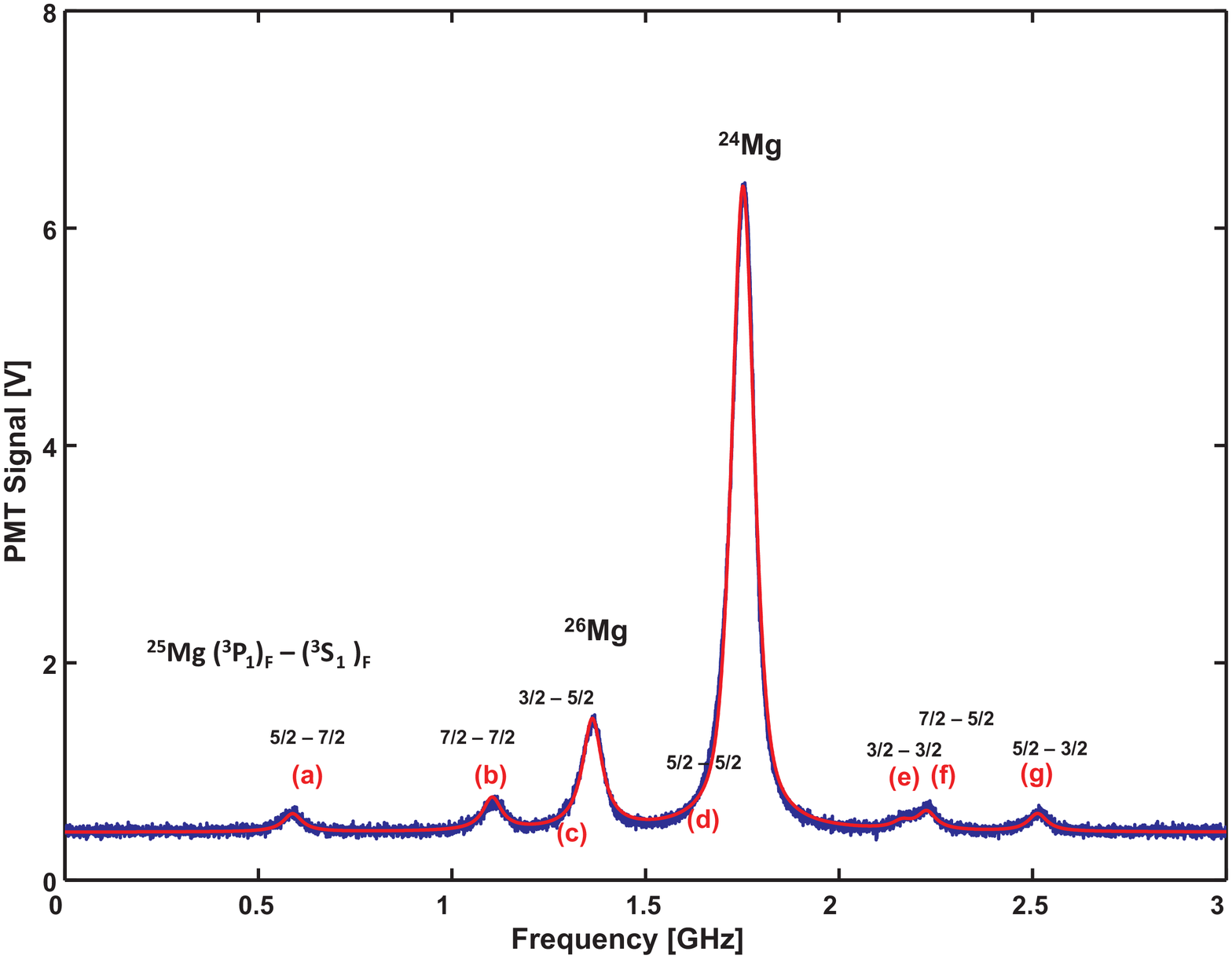}
\caption{(color online) Fluorescence signal from the $^3$P$_1$ -
$^3$S$_1$ transition. The blue curve represents the experimental
data and the red curve is a fit to the data. The hyperfine splitting
of $^{25}$Mg($^3$P$_1$)$_F$ - ($^3$S$_1$)$_F$ are indicated as (a) -
(g).} \label{fig: 3p1}
\end{figure}

\begin{figure}[htb!]
\centering
\includegraphics*[width=1\columnwidth]{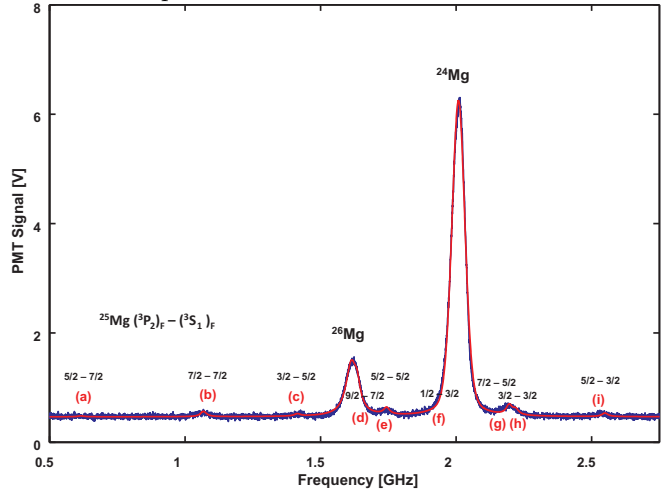}
\caption{(color online) Fluorescence signal from the $^3$P$_2$ -
$^3$S$_1$ transition. The blue curve represents the experimental
data and the red curve is a fit to the data. The hyperfine splitting
of $^{25}$Mg($^3$P$_1$)$_F$ - ($^3$S$_1$)$_F$ are indicated as (a) -
(i).} \label{fig: 3p2}
\end{figure}

Figs. \ref{fig: 3p0}, \ref{fig: 3p1} and \ref{fig: 3p2} show
typical fluorescence spectra as a function of laser frequency for
the $^3$P$_{0,1,2}$ - $^3$S$_1$ transitions. The blue curve is the
raw data and the red curve is a Voigt profile fit to the data. For
$^{24}$Mg we obtain a FWHM linewidth of 55 MHz in agreement with our
beam and detector geometry. From the spectra we clearly identify the
isotope shift and the hyperfine components ($^3$P$_{0,1,2}$)$_F$ -
($^3$S$_1$)$_F$ of $^{25}$Mg. Data for the hyperfine coefficients of
the $^3$P$_{1,2}$ levels are needed to extract the hyperfine
coefficient A($^3$S$_1$) and are taken from \cite{Lurio HFS }. We
extract the hyperfine coefficient A($^3$S$_1$) = (-321.6 $\pm$ 1.5)
MHz. As described in Sec.~\ref{MetCalc} the electric quadrupole
hfs constant B($^3$S$_1$) is significantly smaller than the
resolution of the experiment and is therefore set to zero in the
fitting procedure. Table~\ref{tb:results} summarizes our findings and compares with
previous measurements. Different systematic errors have been
investigated as mentioned in Sec.~\ref{Experimental setup}, all
of which have been determined to be lower than the statistical
error. Errors listed in Table~\ref{tb:results} are pure statistical
errors. Here it can be seen that the value for the hyperfine
coefficient A($^3$S$_1$) and the isotope shifts agree with previous
measurements within the uncertainty.

\begin{table*}[htb!]
\resizebox{1.0\textwidth}{!}{
\begin{tabular}{| c | c | c | c | c | c | c | c |}
\hline & \multicolumn{2}{c|}{} & \multicolumn{2}{c|}{} & \multicolumn{2}{c|}{} & \\

& \multicolumn{2}{c|}{$^3$P$_0$ - $^3$S$_1$} &
\multicolumn{2}{c|}{$^3$P$_1$ - $^3$S$_1$} &
\multicolumn{2}{c|}{$^3$P$_2$ - $^3$S$_1$} &
\\

& \multicolumn{2}{c|}{} & \multicolumn{2}{c|}{} &
\multicolumn{2}{c|}{} & \\ \hline

& & & & & & &\\

& $\Delta^{24-26}$ [MHz] & $\Delta^{24-25}$ [MHz] & $\Delta^{24-26}$ [MHz]
& $\Delta^{24-25}$ [MHz] & $\Delta^{24-26}$ [MHz] & $\Delta^{24-25}$ [MHz] & $A(^3S_1)$ [MHz] \\

& & & & & & & \\ \hline

& & & & & & & \\
Ref. \cite{schwlow isotope} (1949) & 414 $\pm$ 9 & & 366 $\pm$ 45 &
& 414 $\pm$ 12  & &
-322 $\pm$ 6  \\

& & & & & & & \\ \hline

& & & & & & & \\
Ref. \cite{Hallstadius isotope  shift} (1978) & 396 $\pm$ 6  & 210
$\pm$ 36  & 391 $\pm$ 4.5  & 201 $\pm$ 21  & 393 $\pm$
7.5 & 204 $\pm$ 7.5 & -329 $\pm$ 6  \\

& & & & & & & \\ \hline

& & & & & & &\\
Ref. \cite{Gondone isotope} (1990) & 391 $\pm$ 10  & 214 $\pm$ 10 &
393 $\pm$ 10  & 215 $\pm$ 10  & 397 $\pm$ 10 & 217 $\pm$ 10 &
-322 $\pm$ 6  \\

& & & & & & &\\ \hline

& & & & & & &\\
\textbf{This work (2009)} &  \textbf{391.3 $\pm$ 1.7 }  &
\textbf{205.7 $\pm$ 1.5 }  &
 \textbf{390.1 $\pm$ 1.4 }  &  \textbf{209.1 $\pm$ 1.3 }  & \textbf{394.4 $\pm$ 0.8 } & \textbf{205.7 $\pm$ 0.8 } &
\textbf{-321.6 $\pm$ 1.5 } \\
& & & & & & & \\ \hline

\end{tabular}
}
\caption{Measured isotope shift and $^3S_1$ hyperfine structure
constant. The errors listed are the statistical errors.}
\label{tb:results}
\end{table*}

We observe the $^{24}$Mg - $^{26}$Mg shift to be almost constant
which indicates that relativistic isotope shift effects are small or
comparable to the statistical error in our measurement.

Our measured isotope shifts are accounted for by pure mass effect.
In this case the ratio between the $^{24}$Mg - $^{26}$Mg and
$^{24}$Mg - $^{25}$Mg shifts can be expressed as \cite{Hallstadius
isotope  shift}:

\begin{eqnarray}
 \frac{\Delta(^{24}\mbox{Mg}-^{26}\mbox{Mg})}{\Delta(^{24}\mbox{Mg}-^{25}\mbox{Mg})}=
 \frac{26-24}{26 \cdot 24} \cdot \frac{25 \cdot 24}{25-24} =
\frac{25}{13}\simeq
 1.92\;.
\end{eqnarray}

We obtain the ratios: $^{3}$P$_{0}$ $\rightarrow$ $^{3}$S$_{1}$ 1.90
$\pm$ 0.02,$^{3}$P$_{1}$ $\rightarrow$ $^{3}$S$_{1}$ 1.87 $\pm$ 0.01
and $^{3}$P$_{2}$ $\rightarrow$ $^{3}$S$_{1}$ 1.92 $\pm$ 0.01. These
values are consistent with previous results \cite{Hallstadius
isotope  shift,Gondone isotope}.

\section{\label{Conclusion} Conclusion}

In this paper we present improved data for the isotope shift of the
Mg $^3$P$_{0,1,2}$ - $^3$S$_1$ and the hyperfine coefficient
A($^3$S$_1$) for the $^{25}$Mg isotope. We find good agreement
between state of the art many body theory and experimental results.
Experimental values reported here are improved by up to a factor of
ten compared to previous studies.

\begin{acknowledgments}
We would like to acknowledge financial support from the Lundbeck
foundation and the Carlsberg foundation. The authors also gratefully
acknowledge Kjeld Jensen for technical assistance and Julian Berengut
for editing the manuscript.
\end{acknowledgments}

\end{document}